# On entangled photons and local hidden variables

EUGEN MUCHOWSKI[1]

**Abstract:** It is shown that there is no remote action with polarization measurements of photons in singlet state. A model is presented introducing a hidden parameter which determines the polarizer output. This model is able to explain the polarization measurement results with entangled photons. It refutes Bell's Theorem.

## 1. Introduction

'Bell's theorem states that the predictions of quantum theory (for measurements of spin on particles prepared in the singlet state) cannot be accounted for by any local theory.' See Goldstein et al. [1], who give an overview of the state of the discussion about Bell's theorem, which was introduced in 1964 by the Irish physicist. In his paper, "On the Einstein Podolsky Rosen Paradox" [2], Bell developed his famous inequality, which any hidden variable theory describing entangled states has to obey—in contrast to quantum mechanics (QM), which infringes the inequality. This inequality is a relation between the expectation values of measurements taken with different settings of the instruments. Bell stated this theorem in his own words [2]: "In a theory in which parameters are added to quantum mechanics to determine the results of individual measurements, without changing the statistical predictions, there must be a mechanism whereby the setting of one measuring device can influence the reading of another instrument, however remote. Moreover, the signal involved must propagate instantaneously, so that such a theory could not be Lorentz invariant." In other words, nature is not local, as it cannot be described by local realistic models. A general approach to logical Bell inequalities and their relation to contextual models has been detailed in [3]. Bell's paper and many subsequent experiments [4] proving the infringement of Bell's inequality by QM have established the belief by many physicists that nature is non-local [5]. The theory of relativity would not be violated, as no information transport over the quantum channel of entangled photons [6] is possible.

Bell's theorem can be refuted by presenting a counterexample which correctly predicts the expectation values of QM. Many authors have tried this. Some have developed models in which the influence of the measuring apparatus should cause correlations [7]. Others have blamed various loopholes for measuring results that violate Bell's inequality [1]. These have all been ruled out through the results of Delft physicists and others [8-10], who proved that QM correctly describes the measured correlations without any reference to external conditions. One counterexample was presented by Muchowski [11]; however, the derivation was not completely convincing, as it contains ambiguous values for the measurement results. There is also no physical justification for the contextual behaviour of entangled systems. As we will see in this paper, this is given by the indistinguishability of entangled particles. Khrennikov [12] stated that 'Bell nonlocality is a wrong conclusion that the violation of Bell type inequalities implies the existence of mysterious instantaneous influences between distant physical systems'. De Zela [13] demonstrated the possibility of local contextual variables for the Bell theorem. He presented a model 'to exhibit Bell violations in a local-realistic context resorts to inner-product-type correlations, which are mathematically equivalent to Born's rule'.

Looking thoroughly at what Bell has proved, one can see that he only ruled out a specific class of models; namely, those which are not contextual. Non-contextual models cannot describe QM measurement results, as they infringe the Kochen–Specker (KS) theorem. Non-contextuality is defined, by KS, as follows: 'If a QM system possesses a property (value of an observable), then it does so independently of any measurement context.' [14]. Contextual models do not infringe upon the KS theorem. The interdependence between contextuality and indistinguishability, with respect to the KS theorem, has also been discussed in detail [15]. Indistinguishability can be considered not only between identical particles (bosons and fermions), but between information units themselves (i.e., qubits and their possible generalizations). 'This approach suggests information indistinguishability as means for nonlocal realistic entanglement, offering a relatively simple understanding for the limitation of realistic nonlocality in quantum theory' [16].

In this paper, a local realistic but contextual model is presented, where the measurement results are predetermined but the polarization or spin (respectively) of particles in singlet state is not fixed. This is a contextual approach, as verified by the fact that, due to indistinguishability, the polarization or spin of an entangled particle are generally not defined. The model does not imply any restriction on the free choice of the experimenter. The concept of indistinguishability of identical particles could be seen as one of the foundational principles of quantum theory [17]. Quantum systems differ from classic particles, particularly with respect to this effect. The current paper will help to understand experimental results. Such an understanding can only be based on local effects. Although we refer to the singlet state as the basis of the investigation, the model does not make use of the formalism of QM.

[1] E-mail: eugen@muchowski.de

It is definitely not the intention of this paper to question the derivation of Bell's inequality. Bell's proof is mathematically correct. The point is simply that the assumptions of Bell's inequality—namely, the restriction to non-contextual models—do not comprehensively describe conceivable physical reality.

# 2. A hidden variable model for polarization entangled photons

## 2.1. Model overview

The model should reproduce the QM predictions of polarization measurements of entangled photons. In doing so, it would describe measurement results violating Bell's inequality.

Figure 1 shows how entangled photons are generated. Figure 2 shows the experimental arrangement and the co-ordinate systems used. With the system in singlet state, the conditional probability $P_{B\beta|A\alpha}$ that photon 2 passes polarizer PB at $\beta$ if photon 1 passes polarizer PA at $\alpha$ is, according to QM upon projecting the singlet state onto the polarizer's directions

$$P_{B\beta|A\alpha} = \sin^2(\alpha - \beta). \quad (1)$$

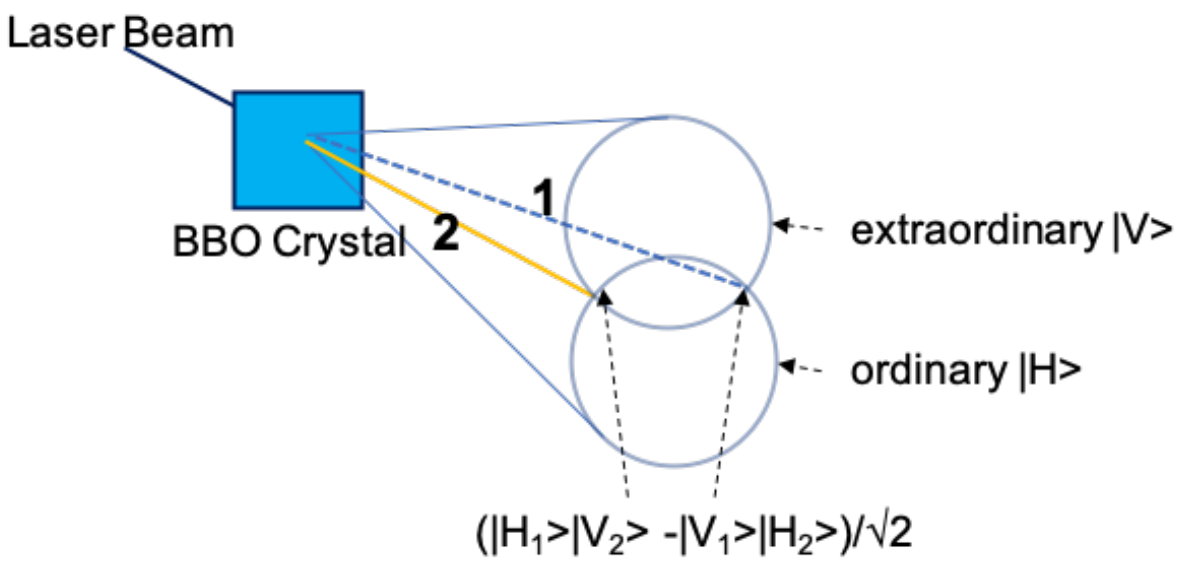

Figure 1: Entangled photons are generated, for example, by parametric fluorescence with a BBO crystal. The ordinary photon beam has polarization 0° and the extraordinary photon beam comes with the polarization 90°. Each photon leaves the source in a cone of light. Both cone shells intersect in beams 1 and 2, which are thus each a mixture of horizontally and vertically polarized photons in equal share. However, due to indistinguishability, their polarization is not defined and the photons are (by superposition) in singlet state $1/\sqrt{2}(|H_1V_2\rangle - |V_1H_2\rangle)$. The Hilbert space for the combined system is $\boldsymbol{H}_{12} = \boldsymbol{H}_1 \otimes \boldsymbol{H}_2$. Normalized base vectors are $|H_1\rangle$ and $|V_1\rangle$ for system 1 at side 1 and $|H_2\rangle$ and $|V_2\rangle$ for system 2 at side 2. $|H_1\rangle$ and $|H_2\rangle$ correspond to the x-axis and $|V_1\rangle$ and $|V_2\rangle$ correspond to the y-axis.

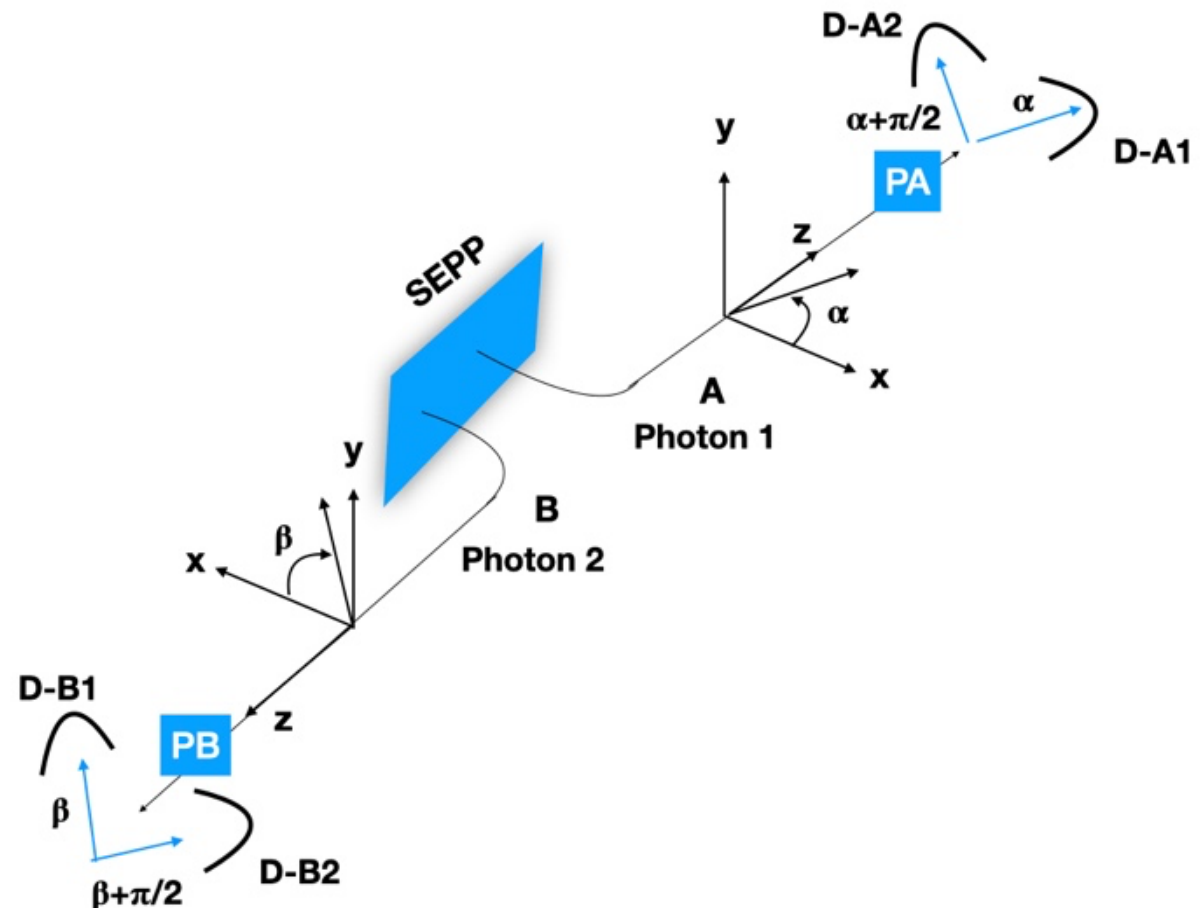

Figure 2: The SEPP (source of entangled photon pairs) emits photons in singlet state propagating in opposite directions towards the adjustable polarizers PA and PB and detectors DA-1 and DA-2 on wing A and DB-1 and DB-2 on wing B. A coincidence measuring device (not seen in the picture) encounters matching events. The polarization angles are defined in the x–y-plane, which is perpendicular to the propagation direction of the photons. The co-ordinate systems are left-handed with the z-axis in propagation direction for each wing, with the x-axis in horizontal and the y-axis in vertical direction.

## 2.2. Model assumptions

With polarization measurements, photons can choose one of two perpendicular polarizer exits. A hidden variable model has to describe which of these two possible exits a photon will take. Five model assumptions are introduced, an overview of which is given below with a description afterwards in italic letters:

| | |
|---|---|
| MA1 | introduces the statistical parameter $\lambda$ which controls the polarizer exit that a photon will take. |
| MA2 | describes how photons from the singlet state are coupled together. |
| MA3 | describes the polarization of a selection of photons from an entangled pair. |
| MA4 | defines how the model parameter δ depends on the chosen context. |
| MA5 | states that photons do not have memory of their previous state after a measurement. |

**Model assumption MA1:**

*The statistical parameter $\lambda$ which is equally distributed between 0 and +1 controls which of the two polarizer exits the photon will take.*

*With polarizer setting $\alpha$ and photon polarization $\varphi$, we define $\delta = \alpha - \varphi$ as the difference between the polarizer setting and the polarization of the photon.*

*The function $A(\delta,\lambda)$ indicates which polarizer exit the photon will take. $A(\delta,\lambda)$ can have values +1 and -1.*

*For $0 \leq \delta < \pi/2$, we define*

$$A(\delta,\lambda) = +1 \quad \text{for} \quad 0 \leq \lambda \leq \cos^2(\delta), \tag{2}$$

*meaning the photon takes polarizer exit $\alpha$ and*

$$A(\delta,\lambda) = -1 \quad \text{for} \ \cos^2(\delta) < \lambda \leq +1, \tag{3}$$

*meaning the photon takes polarizer exit $\alpha + \pi/2$.*

***

The case $\pi/2 \leq \delta < \pi$, is covered referring to the other exit of the polarizer. $\delta < 0$ is covered by reversing the polarizer direction by 180°. Thus, $-\pi < \delta < -\pi/2$ is equivalent to $0 < \delta < \pi/2$ and $-\pi/2 < \delta < 0$ is equivalent to $\pi/2 < \delta < \pi$.

With entangled photons the definition of $\delta$ in equations (2) and (3) depends on the context, see MA4 below.

Figure 3 shows the geometric relationships on which the model is based.

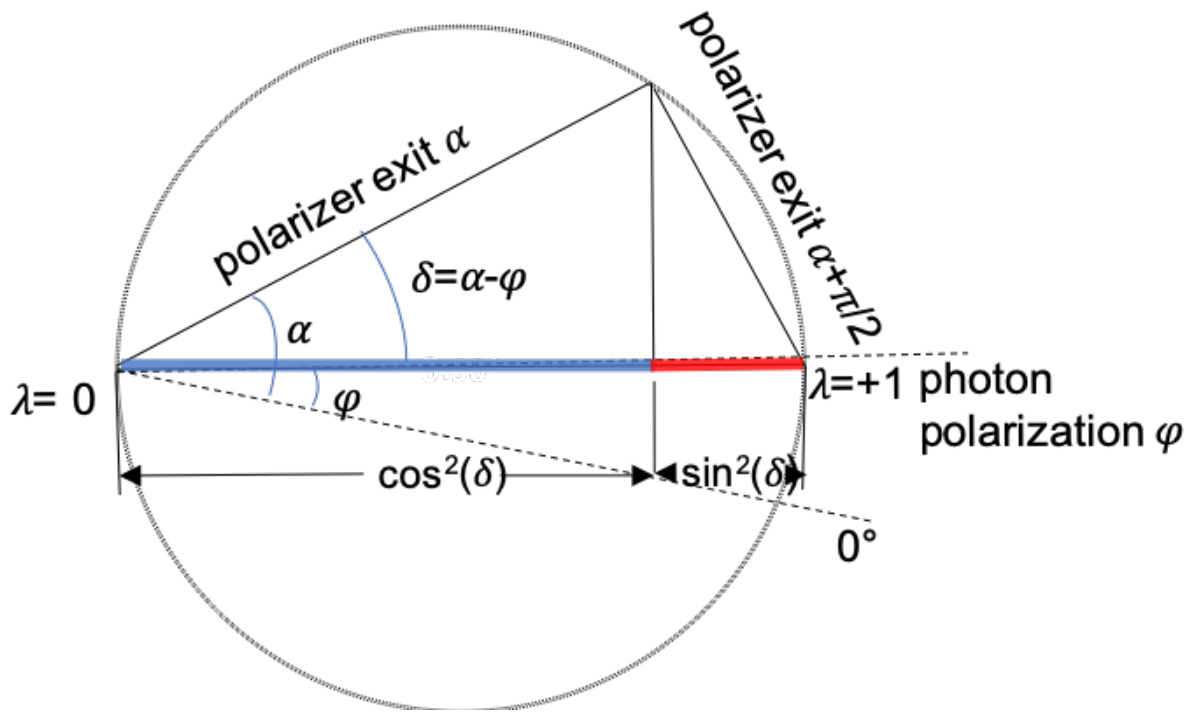

Figure 3: Distribution of polarized photons onto polarizer exits. The polarizer is set to the angle $\alpha/\alpha+90°$ and the photon has a polarization $\varphi$. Thus, the difference is $\delta = \alpha - \varphi$. A parameter $\lambda$ is evenly distributed over the generated photons in the value range $0 \leq \lambda \leq +1$. Photons with $\lambda \leq \cos^2(\delta)$ are assigned to the polarizer exit $\alpha$, while photons with $\lambda > \cos^2(\delta)$ take the exit $\alpha + 90°$.

**Model assumption MA2:**

*Photons of an entangled pair share the same value of the parameter $\lambda$. The rules for the distribution of the photons onto the two exit directions of the polarizer represented by equations (2) and (3) are also valid for the partner photon on wing B.*

***

This reflects the property of the singlet state, with the co-ordinate systems on both wings being of the same handedness with respect to the propagation directions of the photons.

**Model assumption MA3:**

*Selected photons from each wing of the singlet state which would take a polarizer exit $\alpha$ have polarization $\varphi = \alpha$. With a selection other than the initial context, all information about the origin from the initial context is lost.*

***

A selection is an ensemble of photons which all take the same polarizer exit. Photons with polarization $\alpha$ and $\alpha+\pi/2$ come in equal shares, due to symmetry. MA3 accounts for the fact that the polarization of photons from the singlet state is undefined (due to indistinguishability) but changed and re-defined by entanglement. Thus, the photons of a selection cannot be distinguished by their polarization. For a selection of the initial states 0° or 90°, the polarization is not changed as it is already equal to the selected state.

MA3 is a contextual assumption, as the polarization of a selection coincides with the setting of a polarizer. It does not imply any restriction on the free choice of the experimenter, nor the dependence of the hidden variable $\lambda$ on the setting of the measurement instruments. However, it is a local realistic assumption, as it assigns a real value to the physical quantity polarization.

**Model assumption MA4:**

*If the context of entangled photons, given by a polarizer setting, is changed to the setting of the other polarizer, then for this context we replace $\delta$ in all equations with its negative value.*

***

Why MA4 is necessary is explained in the paragraph after equation (9).

**Model assumption MA5:**

*After leaving a polarizer exit, $\lambda$ is indeterminate and uniformly distributed.*

***

MA5 emphasizes the fact that photons do not have memory of their previous polarization after leaving a polarizer. The ensemble of photons covers the full range $0 \leq \lambda \leq +1$ after passing through a polarizer and a photon has polarization $\alpha$ after passing a polarizer with setting $\alpha$.

––––

With the model assumptions defined above, we are going to show, in three steps, that the model:

- predicts the statistical distribution of measurement results with single photons;
- predicts correctly that if a photon leaves a polarizer on one wing, its peer photon certainly leaves a perpendicular set polarizer on the other wing; and
- predicts the expectation value of measurement results of entangled photons.

## 2.3. Predicting measurement results for single photons

Using equation (2), a photon with polarization $\varphi$ is found behind the exit $\alpha$ of a polarizer with probability

$$P_\delta = \int_0^{\cos^2(\delta)} d\lambda = \cos^2(\delta) \qquad (4)$$

where $\delta = \alpha - \varphi$ with $0 \leq \delta < \pi/2$. For the case $\pi/2 \leq \delta < \pi$, we refer to the other exit of the polarizer and have, with $\vartheta^* = \delta - \pi/2$,

$$P_\delta = \int_{\cos^2(\delta^*)}^{1} d\lambda = 1-\cos^2(\vartheta^*) = \cos^2(\vartheta), \text{ as well.} \qquad (4a)$$

With $\delta = \alpha - \varphi$, we obtain the same $P_\delta$ for a photon in state $\cos(\varphi)*|H> + \sin(\varphi)*|V>$ by projection onto $\cos(\alpha)*<H| + \sin(\alpha)*<V|$, according to QM (i.e., Born's rule).

## 2.4. Predicting measurement results for the initial context

In this section, we show that, taking a selection of photons by a polarizer PA set to $\alpha$ on wing A, the thus selected peer photons would leave a polarizer PB exit set to $\alpha+\pi/2$ on wing B with certainty.

Entangled photons are generated by a common source, propagating on wing A with polarization $\varphi_1 = 0°$ and on wing B with polarization $\varphi_2 = 90°$ (or on wing A with the polarization $\varphi_1 = 90°$ and on wing B with the polarization $\varphi_2 = 0°$). The generated photons with polarization 0° and the photons with polarization 90° come in equal shares on either wing; this is the initial context (see Figure 1).

First, we calculate the measurement results for a pair of generated photons: one generated with polarization 0° (photon 1) and another generated with polarization 90° (photon 2).

For photon 1 with $\varphi_1 = 0°$ and an assumed polarizer PA setting of $\alpha$, we would get $\delta = \alpha - \varphi_1 = \alpha$ and, by equation (2),

for $0 \leq \delta < \pi/2$

$$A(\delta,\lambda) = +1 \quad \text{for} \quad 0 \leq \lambda \leq \cos^2(\delta), \qquad (5)$$

meaning that photon 1 takes polarizer PA exit $\alpha$.

By equation (3), we get

$$A(\delta,\lambda) = -1 \quad \text{for} \quad \cos^2(\delta) < \lambda \leq +1, \qquad (6)$$

meaning that photon 1 takes polarizer PA exit $\alpha+\pi/2$; that is, perpendicular to the polarizer PA setting.

Now we proceed to peer photon 2 with polarization $\varphi_2 = 90°$. Defining $B(\delta,\lambda)$ for the measurement results on wing B, we can apply model assumption MA2 describing the common features of the entangled photons in both wings. Here, $\delta$ is again the angle between the polarizer setting and the polarization of the generated photon. Thus, equations (2) and (3) apply, adding 90° to all angles and exchanging $A(\delta,\lambda)$ with $B(\delta,\lambda)$.

For photon 2 with $\varphi_2 = 90°$ and with an assumed polarizer PB setting of $\alpha+\pi/2$, we obtain $\delta = \alpha + \pi/2 - \varphi_2 = \alpha$ and, from equation (2), by exchanging $A(\delta,\lambda)$ with $B(\delta,\lambda)$

*for $0 \leq \delta < \pi/2$*

$$B(\delta,\lambda) = +1 \quad \text{for} \quad 0 \leq \lambda \leq \cos^2(\delta), \qquad (7)$$

meaning that photon 2 would take PB exit $\alpha+\pi/2$; from equation (3), we get

$$B(\delta,\lambda) = -1 \quad \text{for} \quad \cos^2(\delta) < \lambda \leq +1, \qquad (8)$$

meaning that photon 2 would take PB exit $\alpha$. Here, the PB exit $\alpha$ and $\alpha+\pi$ are equivalent.

As entanglement connects photon 1 on wing A with photon 2 on wing B by the same value of the parameter $\lambda$, we obtain from equations (5) and (7) that if photon 1 is detected by PA at $\alpha$, its peer photon 2 is detected with certainty by PB at $\alpha+\pi/2$. From equations (6) and (8), we obtain that if photon 1 is detected by PA at $\alpha+\pi/2$, its peer photon 2 is detected with certainty by PB at $\alpha$.

In the same way, we can calculate measurement results for the pair of generated photon 1 with polarization 90° and generated photon 2 with polarization 0°.

Thus, the model correctly predicts the measurement results with perpendicular polarizers on both wings where all photon 1 which hit PA exit $\alpha$ have peer photon 2 hitting PB at $\alpha+\pi/2$. The reason for this is the common parameter $\lambda$, and not a non-local action, as we have seen.

So far, we have not made use of the contextual assumption MA3 of the model. The derivation, so far, has been local and non-contextual.

## 2.5. Predicting measurement results for an arbitrary context

We now calculate probabilities for an arbitrary setting of the polarizers; that is, having polarizer PA set to $\alpha$ and polarizer PB set to $\beta$. This means changing the selections of the photons. In the initial context (0°/90°), the generated photons with 0° polarization and 90° polarization comprised the selection. Now, the selection is changed and, so, the polarization of the photons —which is defined by model assumption MA3—is also changed.

If PA is set to $\alpha$, all selected photons 1 which take PA exit $\alpha$ have peer photons 2 which would take a polarizer PB' exit $\alpha+\pi/2$, as we have seen above.

The polarizer PB set to $\beta$ selects all photon 2 which would take PB exit $\beta$. By MA3, the polarization of the selected photons 2 is equal to the polarizer setting $\beta$. In order to identify matching events with PA at $\alpha$ and PB at $\beta$ we need to find all photon 2 with polarization $\beta$ which would pass an assumed polarizer PB' at side

B set to $\alpha+\pi/2$. Those photon 2 have a peer photon 1 which would take PA exit $\alpha$ as we have shown above. The probability that photons 2 with polarization $\beta$ would pass PB' at $\alpha + \pi/2$ can be obtained by equation (4,4a), using $\delta = \alpha + \pi/2 - \beta$, yielding

$$P_\delta = \cos^2(\delta) = \cos^2(\alpha + \pi/2 - \beta) = \sin^2(\alpha - \beta), \quad (9)$$

where $\delta$ is the angle between the PB' polarizer setting $\alpha+\pi/2$ and the polarization $\beta$ of photons 2 selected by PB.

It must be assured, that the $\lambda$ values of matching photon pairs are independent on how we look at the experiment, that is which context we choose. In the derivation above we assumed that the context was selected through the selection by PB, so that all photons exiting PB in exit $\beta$ have the polarization $\beta$. One could also look at the experiment from the other side and define the context through PA. In this case, we have to exchange $\beta$ with $\alpha$ and $\alpha + \pi/2$ with $\beta - \pi/2$ in the derivation of equation (9). Thus, we would get $\delta = \beta - \alpha - \pi/2$ with an opposite sign compared to above with the effect that matching pairs would then have $\lambda$ values in the range $\sin^2(\delta) < \lambda \leq +1$ instead of $0 \leq \lambda \leq \cos^2(\delta)$; (see remarks below MA1 and equations (4, 4a)). This is prevented by model assumption MA4.

The expectation value for joint measurement with photon 1 detected behind detector PA at $\alpha$ and peer photon 2 detected behind detector PB at $\beta$ is

$$E(\alpha,\beta) = \Sigma(A*B*P_{A,\alpha} * P_{B,\beta|A,\alpha} \quad \text{for A,B=+1,-1}), \quad (10)$$

where $P_{A,\alpha}$ is the unconditional probability to detect (A = 1) or not detect (A = -1) photon 1 at $\alpha$ and where $P_{B,\beta|A,\alpha}$ is the conditional probability for photon 2 to have the outcome B (+1 for passing or -1 for not passing) at PB set to $\beta$ if photon 1 has outcome A (+1 for passing or -1 for not passing) at PA set to $\alpha$.

With $P_\delta$ from equation (9), we get, in particular, that:

$$P_{+1,\beta|+1,\alpha} = P_{-1,\beta|-1,\alpha} = P_\delta \text{ and}$$

$$P_{+1,\beta|-1,\alpha} = P_{-1,\beta|+1,\alpha} = 1 - P_\delta. \quad (11)$$

Generated photons of polarization 0° and 90° contribute to the probability $P_{1,\alpha}$ to find a photon 1 at $\alpha$, according to equation (4,4a), with fractions $\frac{1}{2}\cos^2(\alpha)$ and $\frac{1}{2}\cos^2(\alpha-\pi/2)$, respectively, taking into account that the generated photons of 0° and 90° contribute in equal shares to the total stream of photons on either wing. Thus,

$$P_{1,\alpha} = \tfrac{1}{2}(\cos^2(\alpha) + \cos^2(\alpha-\pi/2)) = \tfrac{1}{2} = P_{-1,\alpha}. \quad (12)$$

With the above definitions, we get, from equations (9)–(11):

$$E(\alpha,\beta) = \tfrac{1}{2}(1* P_\delta -1*(1- P_\delta) -1*(1- P_\delta) +1* P_\delta) =$$

$$= P_\delta - (1- P_\delta) = \sin^2(\alpha - \beta) - \cos^2(\alpha - \beta) = -\cos(2(\alpha - \beta)), \quad (13)$$

in accordance with QM.

As the expectation value $E(\alpha,\beta)$ in equation (13) exactly matches the predictions of quantum physics, it also violates Bell's inequality.

## 2.6. Extending the model to spin ½ particles

The model also applies to spin ½ particles, by simply replacing every angle with its half in equations (2) and (3) and Figure 3, as well as in all subsequent derivations, yielding

$$E(\alpha,\beta) = -\cos(\alpha - \beta), \quad \text{in accordance with QM.} \quad (14)$$

Perpendicular polarizer exits correspond to opposite instrument exits. Figure 4 shows the geometric relationships on which the model is based for spin ½ particles.

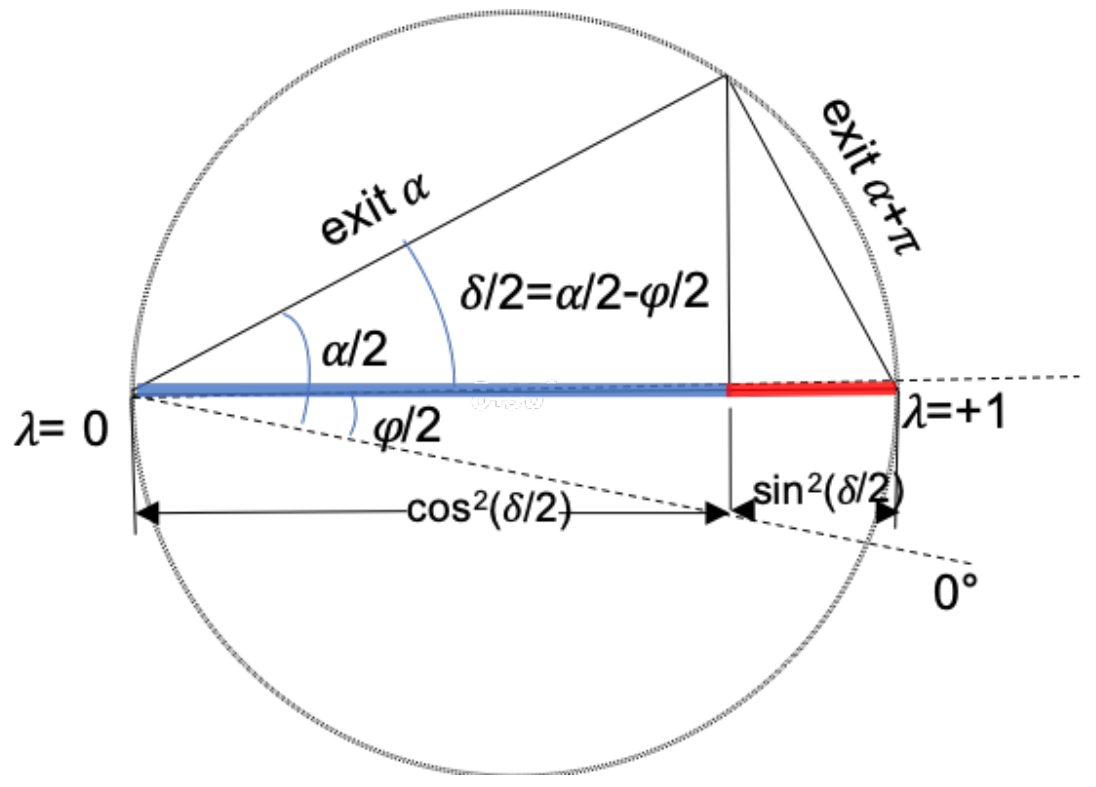

Figure 4: Geometric derivation of a deterministic distribution of spin ½ particles onto instrument exits. The instrument is set to the angle $\alpha/\alpha+180°$. The generated particle has a spin direction $\varphi$. The difference is $\delta = \alpha - \varphi$. The parameter $\lambda$ is evenly distributed over the generated particles in the value range of $0 \leq \lambda \leq +1$. Particles with $\lambda \leq \cos^2(\delta/2)$ are assigned to the instrument exit $\alpha$, while particles with $\lambda > \cos^2(\delta/2)$ take the exit $\alpha+180°$.

# 3. Results, Discussion, and Conclusion

A local contextual model was presented which correctly meets the predictions of QM for polarization measurements with photons in singlet state, as well as for spin measurements with spin 1/2 particles in singlet state, thus refuting Bell's theorem.

The rule determining which instrument exit a particle will take is the same for both wings. Dependencies between the particles in either wing originate from the shared parameter $\lambda$, and not from a non-local influence of particle 1 upon particle 2. With the same rules acting upon the generated particles in both wings, the measurement results are correlated without non-local effects. The model is also valid for single particles.

Measurement values at one wing are determined independently of the setting of the instrument at the other wing. However, the polarization or the spin respectively of particles in singlet state are not fixed. That is a contextual effect that comes into play as the spin or polarization of entangled particles are undefined. The model says that the spin/polarization of a selection of particles is changed by entanglement due to indistinguishability. However, this effect is local, as we have seen. The purpose of the model was to show that a local contextual model which correctly predicts expectation values in accordance with QM is possible. The model's predictions violate Bell's inequality, thus refuting Bell's theorem stating that local realistic models are impossible. Mathematically there is no error in Bell's derivation. But physically he did not take into account the effect of indistinguishable particles. Hence, the premises of Bell's derivation are not complete enough for a description of the quantum reality.

After Bell's theorem is refuted, we can no longer conclude that nature is non-local. This is no more a necessary consequence of QM infringing upon Bell's theorem. Experimental results with spin or polarization measurements can be explained without assuming non-local effects. If there is no basis for non-local effects anymore, some conclusions can be drawn; regardless of whether or not the model presented can replace QM in every detail. It is assumed that the observations on particles from an entangled pair are generally also valid for single particles.

1. Measured values are not generated upon measurement, they already exist beforehand. Otherwise, a strong correlation between the outcomes of measurements at different sides would demand non-local effects.
2. This supports Einstein's view of the meaning of the wave function as a description of an ensemble [18]. Thus, quantum mechanics does not violate the principle of causality; at least for spin measurements.
3. The concept of superposition, which implies the simultaneous existence of incompatible physical states, is in question. If measured values exist beforehand, mutually exclusive values cannot exist simultaneously.

As a consequence, the concept of a quantum computer also comes into question, as it relies upon the assumption that a quantum system bears simultaneous information about two mutually exclusive outcomes [19]. As this assumption is no longer tenable, the diversity of the solution of a quantum computer is considerably restricted.